\documentclass{czjphys}         
\usepackage{epsfig}
\begin{document}
\title{Higher Twist Effects in Hadronic B Decays}
\authori{Kwei-Chou Yang\footnote{Email address:
kcyang@cycu.edu.tw}}
\addressi{Department of Physics, Chung Yuan Christian
University, Chung-Li, Taiwan 320, Republic of China}
\authorii{}
\addressii{}
\authoriii{}    \addressiii{}
\authoriv{}     \addressiv{}
\authorv{}      \addressv{}
\authorvi{}     \addressvi{}
 \headauthor{Kwei-Chou Yang}
 \headtitle{Higher Twist Effects in
 Hadronic B Decays}
 \lastevenhead{Kwei-Chou Yang: Higher Twist Effects in
 Hadronic B Decays ldots}
\pacs{13.20.Hw,12.39.St,12.38.Bx} \keywords{QCD factorization,
light-cone distribution amplitude}
\refnum{A}
\daterec{30 June 2003}   
\issuenumber{x}  \year{2003} \setcounter{page}{1}
\maketitle
\begin{abstract}
Within the framework of QCD factorization, we discuss various
important corrections arising from higher twist distribution
amplitudes of mesons in the hadronic $B$ decays.
\end{abstract}

\section{Factorization with QCD improvement}
Consider the charmless $B\to M_1 M_2$ decays with $M_2$ being
emitted, as shown in Fig.~\ref{fig:QCDF}. The decays can be
studied by means of the QCD factorization (QCDF)
approach~\cite{BBNS1}. Since the energies of final state mesons
$M_1, M_2$ are order of $m_B/2$, the soft corrections between the
two mesons will therefore decouple in order of $\Lambda_{\rm
QCD}/m_b$. Only hard interactions between $(BM_1)$ and $M_2$
survive in the heavy $b$ quark mass limit, $m_b\to\infty$, and
soft effects are confined to $(BM_1)$ system. In the QCDF
approach, the transition matrix element of the 4-quark operator
$O_i$, depicted in Fig.~\ref{fig:QCDF}, is given by
\begin{eqnarray} \langle M_1M_2|O_i|B\rangle
&&=F^{BM_1}(m_2^2)\int^1_0 du T^I(u)\Phi_{M_2}(u)\nonumber\\
&&+\int^1_0 d\xi\,du\,dv
\,T^{II}(\xi,u,v)\Phi_B(\xi)\Phi_{M_1}(v)\Phi_{M_2}(u),
\end{eqnarray}
 \noindent
where $T^I,~T^{II}$ are hard scattering functions, $F^{BM_1}$ the
$B\to M_1$ transition form factor, $\Phi_{M_{1,2}}$ the light-cone
distribution amplitudes (LCDAs) of the final state mesons.
%
\bfg[t]                     
\bc                         
\centerline{\epsfig{file=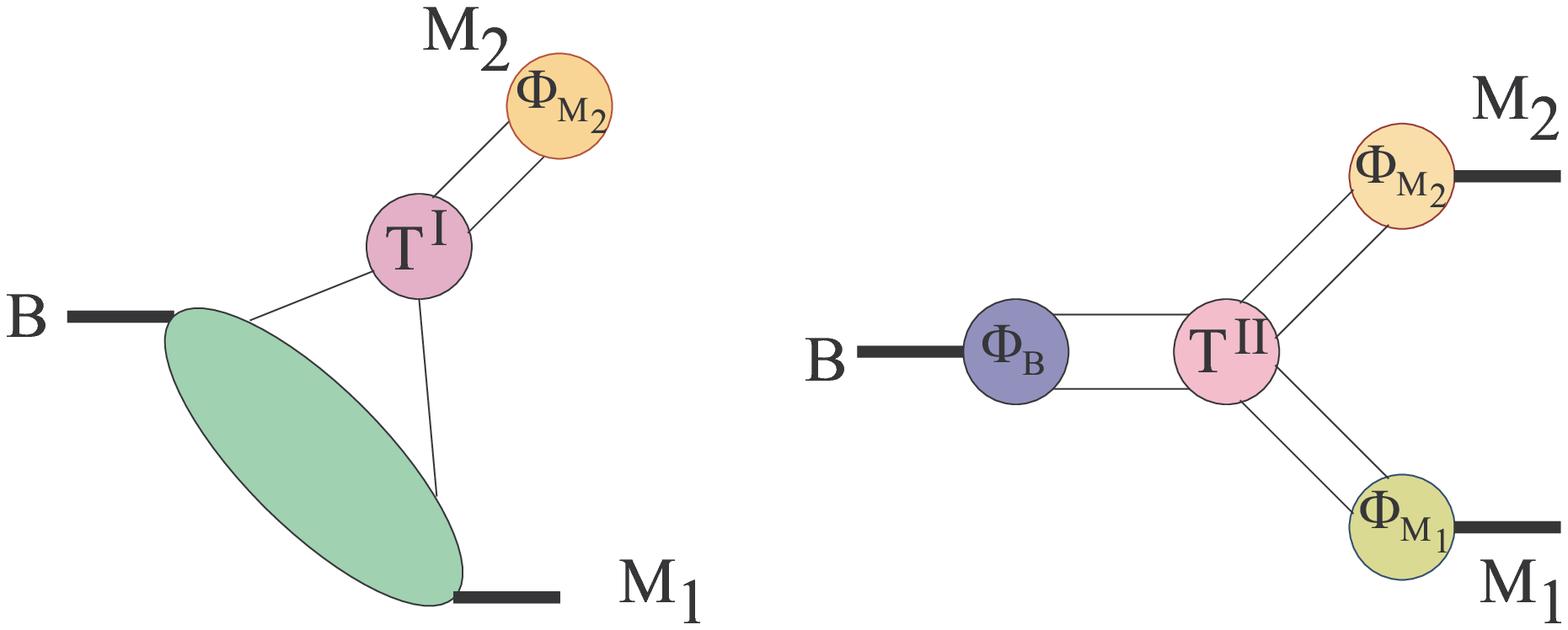,width=70mm,clip=50mm},\epsfig{file=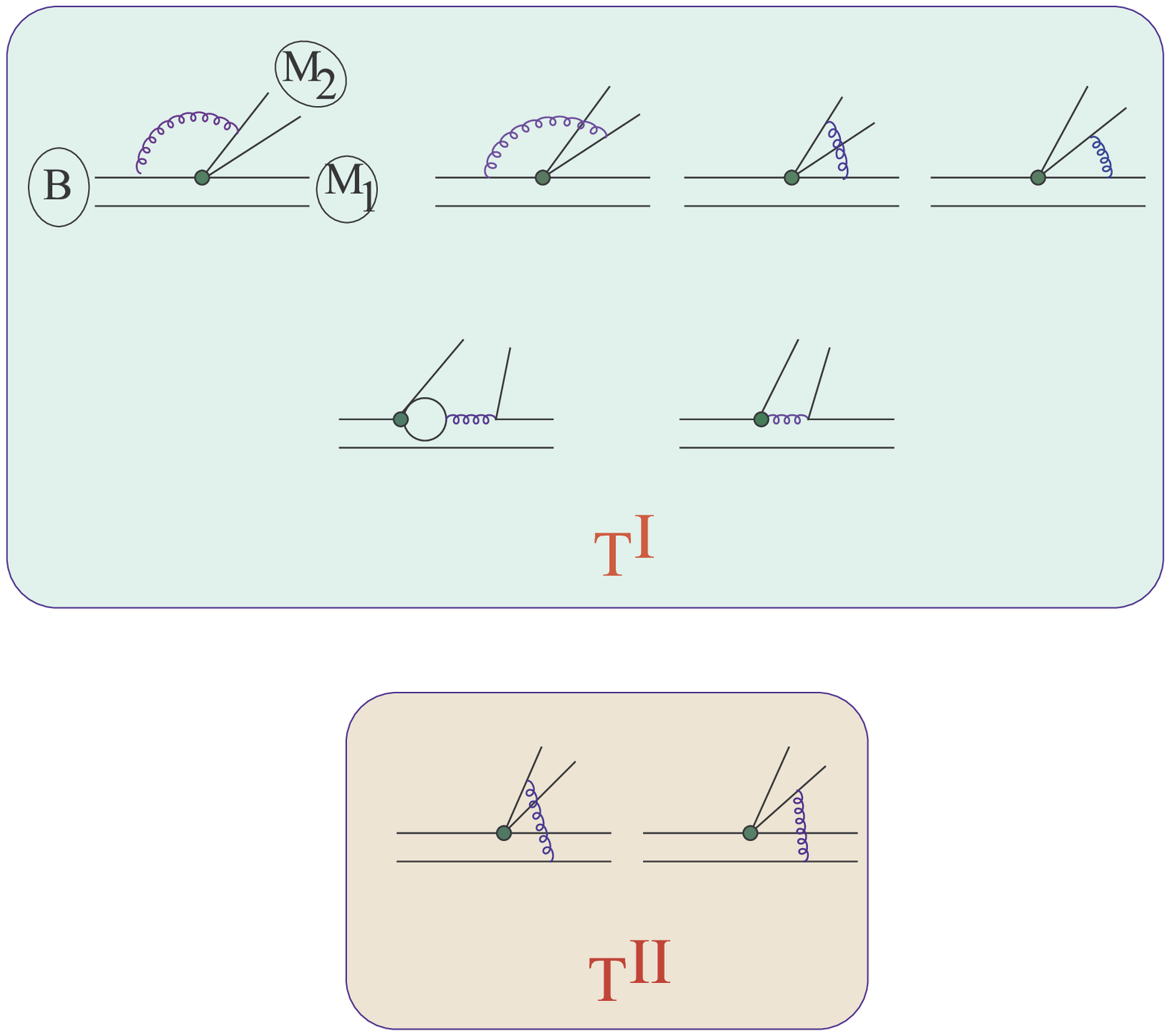,width=50mm,clip=50mm}}
\ec                      
\vspace{-6mm} \caption{Pictorial representation of QCD
factorization formula in $B$ decays.}\label{fig:QCDF}
\efg                        

\section{Light-cone distribution amplitudes of light mesons}
The nonlocal quarks (and gluon(s)) sandwiched between the vacuum
and the final state meson can be expressed in terms of a set of
LCDAs. For the processes of $B\to VP$, where $V$ and $P$ denote
the vector and pseudoscalar mesons, respectively, we find that the
weak annihilation diagrams induced by $(S-P)(S+P)$ penguin
operators, which are formally power-suppressed by order
$(\Lambda_{\rm QCD}/m_b)^2$, are chirally and logarithmically
enhanced owing to the non-vanishing end-point behavior of the
twist-3 LCDA of the pseudoscalar meson $P$~\cite{Cheng:2000hv}.
The two-parton LCDAs of the light pseudoscalar meson of interest
are given by \cite{Cheng:2000kt,Cheng:2000hv}
 \bea \langle P(p)|\bar
q_\alpha(x)q'_\beta(0)|0\rangle&=&{if_P\over 4}\int^1_0
du\,e^{iup\cdot
x}\nonumber\\
&&\times \Bigg[p\!\!\!\!/\gamma_5\Phi(u) -
\mu_\chi\gamma_5\Big(\Phi_p(u)-{1\over 6}\sigma_{\mu\nu}p^\mu
x^\nu\Phi_\sigma(u)\Big)\Bigg]_{\beta\alpha}\,,
 \eea
where $\Phi$ is the leading twist (twist-2) LCDA and $\Phi^P_p$
and $\Phi^P_\sigma$ are of twist-3. The above LCDAs are defined by
 \bea
  \langle P(p)|\bar q_1(0)\gamma_\mu\gamma_5 q_2(x)|0\rangle &=&-if_P
p_\mu\int^1_0 d\bar\eta\,e^{i\bar\eta P\cdot x}\phi^K(\bar\eta),\nonumber\\
\langle P(p)|\bar q_1(0)i\gamma_5 q_2(x)|0\rangle &=& f_P
\mu_\chi^P\int^1_0
d\bar\eta\,e^{i\bar\eta p\cdot x}\Phi_p^P(\bar\eta), \nonumber\\
\langle P(p)|\bar q_1(0)\sigma_{\mu\nu}\gamma_5 q_2(x)|0\rangle
&=& -{i\over 6}f_P\mu_\chi^P\left[1-\left({m_1+m_2\over
m_P}\right)^2\right]\nonumber\\
&\times&(p_\mu x_\nu-p_\nu x_\mu) \int^1_0 d\bar\eta
\,e^{i\bar\eta p\cdot x}\Phi_\sigma^P(\bar \eta),
\label{eq:tensor}
 \eea
where $\mu_\chi^P=m_P^2/(m_1+m_2)$, with $m_{1,2}$ being the
current quark masses of $q_{1,2}$, and the asymptotic forms of
LCDAs are: \bea &&\Phi^{P}(x)=6x(1-x),\nonumber \\
&&\Phi^P_p(x)=1, \quad \Phi^P_\sigma(x)=6x(1-x).
\label{eq:t3da}\eea

\section{Higher twist effects in $B\to \phi K$}
CLEO, BaBar and Belle recently reported the
results~\cite{Briere,babar_omegaK,belle_omegaK}
  \bea
 {\cal B}(B^\pm\to\phi K^\pm)&=&\cases{
(5.5^{+2.1}_{-1.8}\pm0.6)\times 10^{-6} & CLEO, \cr
(10.0\pm0.9\pm0.5)\times 10^{-6} & BaBar, \cr
(14.6\pm3.0^{+2.8}_{-2.0})\times 10^{-6} & Belle,}\nonumber\\
 {\cal B}(B^0\to\phi K^0)&=&\cases{ (5.4^{+3.7}_{-2.7}\pm0.7)&
CLEO, \cr (7.6\pm1.3\pm0.5)& BaBar, \cr (13.0^{+6.1}_{-5.2}\pm2.6)
& Belle.}
 \eea
In absence of annihilation contributions, the resulting branching
ratios are $Br(B^-\to\phi K^-)=(3.8\pm0.6)\times 10^{-6}$ and
$Br(B^0\to\phi K^0$)= $(3.6\pm0.6)\times 10^{-6}$, which are small
compared with the data. The relevant weak annihilation diagrams
for $\phi K$ modes, which are penguin-dominated processes, are
depicted in Fig.~\ref{fig:ann}. The annihilation contributions
induced by $(S-P)(S+P)$ penguin operators are not subject to
helicity suppression and could be sizable. As shown in
\cite{Cheng:2000hv}, these annihilation contributions, which are
formally power-suppressed by order of $(\Lambda_{\rm QCD}/m_b)^2$,
are chirally and logarithmically enhanced. The logarithmical
divergence (or enhancement) is owing to the non-vanishing
end-point behavior of the twist-3 LCDA of the kaon. The
annihilation amplitude for $\phi K$ modes is approximately given
by
%
\bfg[t]                     
\bc                         
\epsfig{file=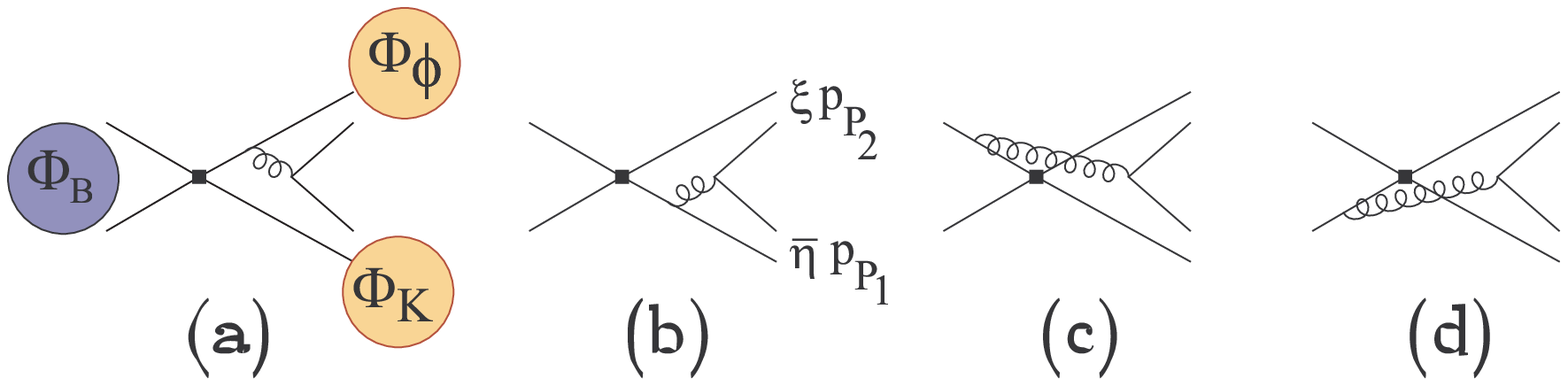,width=100mm,clip=50mm}
\ec                      
\vspace{-2mm} \caption{Annihilation diagrams for $B\to \phi K$
decays.}\label{fig:ann}
\efg                        

\bea {\cal A}_{\rm ann}&\simeq& -\frac{G_F}{\sqrt{2}} V_{tb}
V_{ts}^* f_B f_K f_\phi
  b_3(\phi, K),
  \eea
where \be b_3(\phi, K)\approx \frac{C_F}{N_c}
(c_6(\mu_h)+c_5(\mu_h)/N_c)\times 6\pi\alpha_s
\frac{2\mu_\chi^K}{m_B}(2X_A^2-X_A), \ee
 with $X_A$ being usually parametrized as
 $X_A=(1+\rho_A)\ln(m_B/\Lambda_\chi)$.
Note that the gluon propagator in the annihilation diagrams, as
shown in Fig.~\ref{fig:ann}, is not as hard as in the vertex
diagrams. Since the virtual gluon's momentum squared there is
$(\bar\eta p_{P_1}+\xi p_{P_2})^2\approx \bar\eta \xi m_B^2 \sim
1$~GeV$^2$, where $\bar\eta m_B\sim \Lambda_\chi$ and $\xi\sim
0.5$, the relevant scale for the annihilation topology should be
$\mu_h\sim 1$~GeV.

By comparing calculation results with the data, we therefore know
that annihilation contributions are not negligible and could give
50\% corrections to the decay amplitudes of $\phi K$. Moreover,
because the annihilation amplitudes give constructive
contributions to $\phi K$ modes, they need to have the large real
part, $Re(\rho_A)\geq 0.7$ for adopting $f_B=180$~MeV. A totally
different conclusion has been made in \cite{Chen:2001pr}. Their
calculation gives almost imaginary annihilation contributions to
the decay amplitudes. Nevertheless, their resulting branching
ratios are instead enhanced by matching the full theory to
effective theory at the confinement scale, where the Wilson
coefficients become much larger. Using the QCDF approach, we
obtain $\phi K^*/ \phi K \sim 1$~\cite{Cheng:2000hv,Cheng:2001aa},
while in the pQCD study \cite{Chen:2001pr,Chen:2002pz} the ratio
is $\sim 1.5$. The result should be testable in the near future
measurements.

\section{Higher twist effects in $B\to V V$} The $B\to VV$
amplitude consists of three independent Lorentz scalars:
 \bea
{\cal A}(B(p)\to V_1(\varepsilon_1,p_1)V_2(\varepsilon_2,p_2))
\propto
\varepsilon_1^{*\mu}\varepsilon_2^{*\nu}(ag_{\mu\nu}+bp_\mu
p_\nu+ic\epsilon_{\mu\nu\alpha\beta}p_1^\alpha
p_2^\beta),\nonumber \label{amp}
 \eea
 where $c$ corresponds to the $p$-wave amplitude, and $c,d$ are
related to the mixture of $s$- and $d$-wave amplitudes. The three
helicity amplitudes are given by
 \bea H_{00} &=& {1\over
2m_1m_2}\left[ (m_B^2-m_1^2-m_2^2)a+2m_B^2p_c^2b\right],  \nonumber \\
H_{\pm\pm} &=& a\mp m_Bp_c c,\nonumber
 \eea where $p_c$
is the c.m. momentum of the vector meson in the $B$ rest frame and
$m_{1,2}$ is the mass of the vector meson $V_{1,2}$. Take $B\to
\phi K^*$ as an example, which is shown in Fig.~\ref{fig:vv}. When
compared with $H_{00}$, to occur in the $H_{--}$ helicity
amplitude, the spin of the $\bar s$ in the emitted vector meson
$\phi$ has to be flipped. Therefore, the amplitude $H_{--}$ is
suppressed by a factor of $m_\phi/m_B$. The $H_{++}$ amplitude is
subject to a further spin flip and therefore is suppressed by
$(m_\phi/m_B)\times (m_{K^*}/m_B)$. It is thus expected that
$|H_{00}|^2 \gg |H_{--}|^2 \gg |H_{++}|^2$.
%
\bfg[t]                     
\bc                         
\epsfig{file=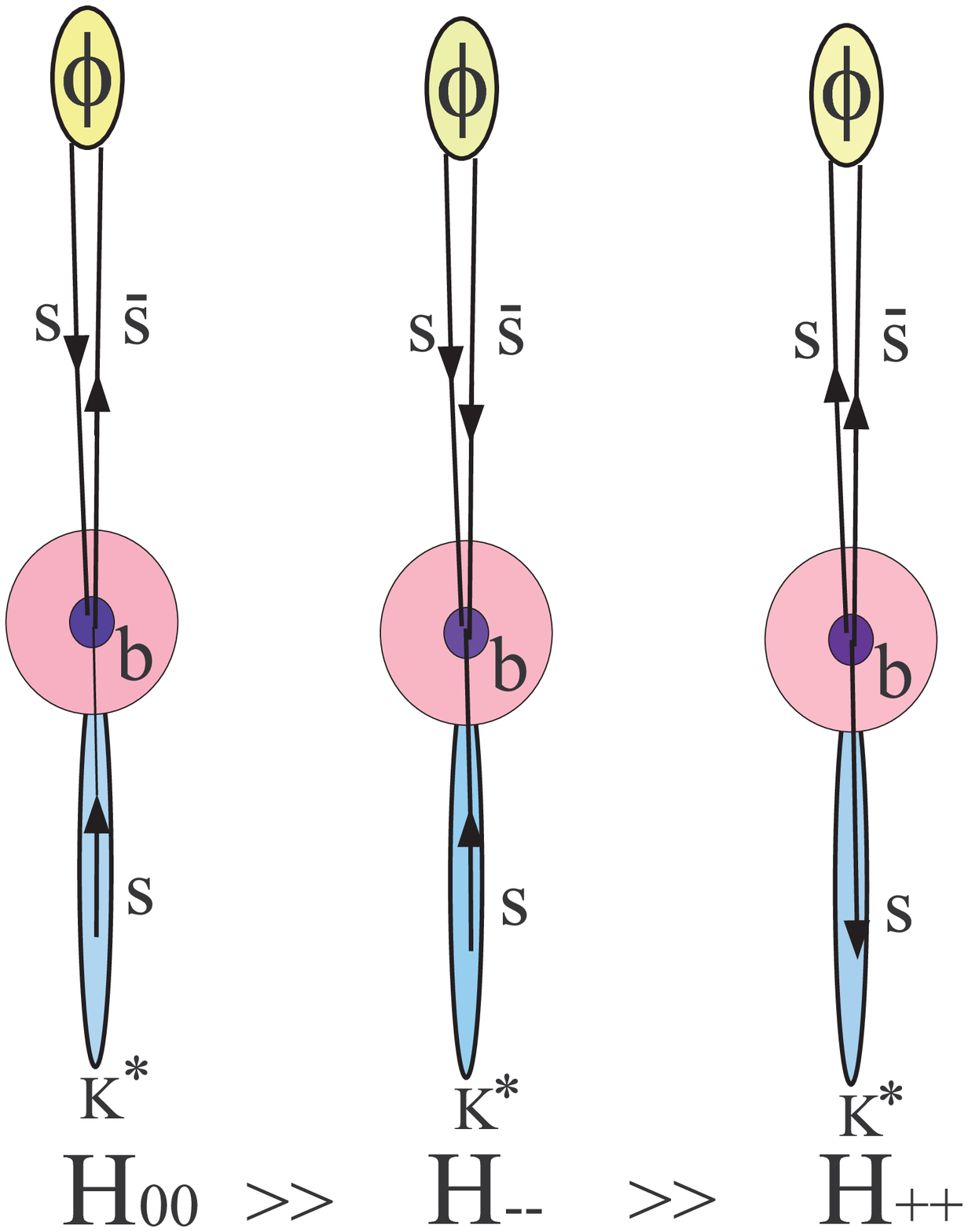,width=40mm,clip=50mm}
\ec                      
\vspace{-2mm} \caption{The directions of quark spins in $B\to \phi
K^*$ helicity amplitudes. Here the decays are assumed to happen
via a $(V-A)(V-A)$ 4-quark operator.}\label{fig:vv} \efg The QCDF
results indicate that the nonfactorizable correction to each
helicity amplitude is not the same; the effective Wilson
coefficients $a_i$ vary for different helicity amplitudes. The
leading-twist nonfactorizable corrections to the transversely
polarized amplitudes vanish in the chiral limit and hence it is
necessary to take into account twist-3 DAs $g_\perp^{(v,a)}$ of
the vector meson in order to have renormalization scale and scheme
independent predictions. Because the $(S-P)(S+P)$ penguin
contributions to the $W$-emission amplitudes are absent and
annihilation is always suppressed by helicity mismatch,
tree-dominated decays tend to have larger branching ratios than
the penguin-dominated ones~\cite{Cheng:2001aa}.

\section{Higher twist effects in $B\to J/\psi K$}
%
\bfg[t]                     
\bc                         
\epsfig{figure=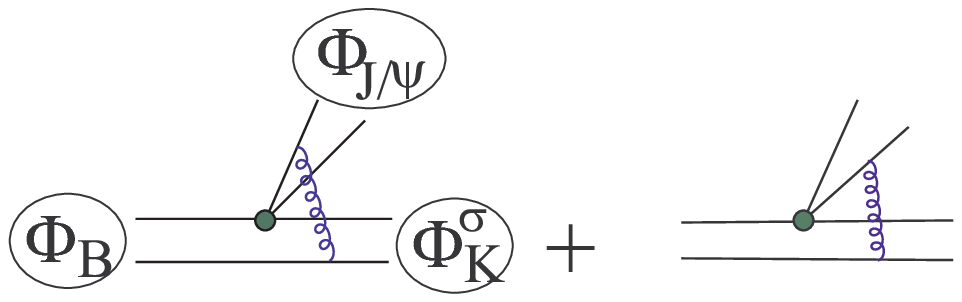,height=25mm}
\ec                         
\caption{Spectator corrections to $B\to J/\psi
K$.}\label{fig:jpsik}
\efg                        
The $B\to J/\psi K^{(*)}$
modes~\cite{Cheng:2000kt,Cheng:2001ez,Cheng:2001cs} are of great
interest because they are only few of color suppressed modes in
hadronic $B$ decays that have been measured. They receive large
nonfactorizable corrections. Under factorization, the $B\to J/\psi
K$ decay amplitude reads
 \be
{\cal A}(B\to J/\psi K)\cong\,{G_F\over\sqrt{2}}V_{cb}V_{cs}^* a_2
f_{J/\psi} m_{J/\psi} F_1^{BK}(m^2_{J/\psi})(2\varepsilon^*\cdot
p_B).
 \ee
$|a_2|$ can be extracted from the data and its value is
$|a_2|\simeq 0.27\pm 0.04$, where the error depends on the form
factor model of $F_1^{BK}$. The QCDF result for $a_2$ is
 \bea a_2= c_2+{c_1\over N_c}+{\alpha_s\over
4\pi}\,{C_F\over N_c} c_1\left[-18-12\ln{\mu\over m_b}+f_I+
\frac{F_0^{BK}(m^2_{J/\psi})}{F_1^{BK}(m^2_{J/\psi})}g_I+f_{II}^2+
f_{II}^3\right],
 \eea
where $f_I, g_I$ arise from the vertex corrections, $f_{II}^2$
from the twist-2 hard spectator interaction, and $f_{II}^3$ from
the twist-3 hard spectator interaction. In Fig.~\ref{fig:jpsik} we
plot the spectator corrections to the decay amplitude of $B\to
J/\psi K$. To leading-twist order, i.e. neglecting $f_{II}^3$, we
have $a_2(J/\psi K)\sim 0.15$ which is too small when compared
with the data. The contribution of two-parton kaon LCDAs of
twist-3 to the spectator diagrams is
 \bea f^3_{II}&&=\left( 2\mu_\chi^K\over m_B \right){4\pi^2\over
N_c}\,{f_K
f_B\over F_1^{BK}(m^2_{J/\psi})m_B^2}{1\over (1-m_{J/\Psi}^2/m_B^2)^3}\nonumber\\
&&\times\int^1_0 dx dy dz\,{\Phi^B(x)\over
x}{\Phi^{J/\psi}(y)\over y}\,{\Phi^K_\sigma(z) \over
6z^2},\label{eq:t3spectator}
 \eea where $\Phi^K_\sigma$ is the
two-parton kaon LCDA of twist-3 which has been shown in
Eqs.~(\ref{eq:tensor}) and (\ref{eq:t3da}). In the above equation,
the integral of $\Phi^K_\sigma$ is divergent. However, it is known
that the collinear expansion cannot be correct in the end point
region owing to the non-zero transverse momentum $\langle
k_\perp\rangle$ of the quark. Thus we parametrize the integral as
 \be\int_0^1{dz\over z}\,{\Phi^{K}_\sigma(z)\over
6 z}\sim \int_0^1{dz\over z+\langle 2k_\perp
\rangle/m_b}-1\simeq\ln (m_B/\Lambda_\chi)(1+\rho_H)-1 \ee
 To account for the experimental value of $|a_2|$, the parameter
$\rho_H$ should be chosen as large as $\sim
1.5$~\cite{Cheng:2000kt}. It implies that $a_2(J/\psi K)$ may be
largely enhanced by the nonfactorizable spectator interactions
arising from the twist-3 kaon LCDA $\Phi^K_\sigma$, which are
formally power-suppressed but chirally and logarithmically
enhanced~\cite{Cheng:2000kt}.

Nevertheless, since the contribution of the twist-3 kaon LCDA
$\Phi^K_\sigma$ to the spectator diagram is end-point divergent in
the collinear expansion, the vertex of the gluon and spectator
quark should be considered to be inside the kaon wave function.
I.e., the kaon itself is at a three-parton Fock state. Instead of
considering the contribution of the two-parton kaon LCDA of
twist-3 to spectator diagrams, we calculate the subleading
corrections from the three-parton LCDAs of the kaon and get $a_2=
0.27+0.05i$~\cite{yang1} which is well consistent with the data.
The result also resolves the long-standing sign ambiguity of $a_2$
which turns out to be positive for its real part.

\section{Can we understand why $K\omega/\pi\omega\sim 1 ?$}
The history of searching for the $B^- \to \omega K^-$ rate is very
interesting. The $\omega K^-$ mode was first reported by CLEO in
1998~\cite{cleo-prl-81} with a large branching ratio $\sim
15\times 10^{-6}$, but disappeared soon after analyzing a larger
data set~\cite{cleo-prl-85} which was confirmed by the later BABAR
measurement~\cite{babar-prl-87}. However, recently Belle observed
a large $\omega K^-$ rate, $(9.2^{+2.6}_{-2.3}\pm 1.0)\times
10^{-6}$, and $\omega K^-/ \omega\pi^- \sim 2$~\cite{belle}. The
non-small $\omega K^-$ rate is also shown in the newly BABAR
data~\cite{babar_omegaK} with $\omega K^- \sim \omega\overline K^0
\sim \omega\pi^- \sim 5\times 10^{-6}$. From the theoretical point
of view, large $\omega K$ rates are hard to understand. The ratio
$\overline K^0 \omega/\pi^- \omega$ reads
 \bea
\frac{\overline K^0\omega}{\pi^-\omega} \approx
\left\vert\frac{V_{cb}}{V_{ub}}\right\vert^2
\left(\frac{f_K}{f_\pi}\right)^2\left\vert {a_4-a_6 r_\chi^K+
(F^{BK}_1 f_\pi)/(F^{B\pi}_1 f_K)r_1 a_9/2 +f_B f_K b_3(K,\omega)
\over a_1+r_1 a_2} \right\vert^2,
 \eea where $r_1 = f_\omega
F_1^{B\pi} /f_{\pi} A_0^{B\omega}$, the chirally enhanced factor
$r^K_\chi= \frac{2m_K^2}{m_b(m_s+m_u)}$ with $m_{s,u}$ being the
current quark masses, and $b_3(K,\omega)$ is the annihilation
contribution of $K\omega$ modes~\cite{Du:2002up}. The
$\pi^-\omega$ rate weakly depends on the annihilation effects.
Without annihilation, since the $a_4$ and $a_6 r_\chi^K$ terms are
opposite in sign in the $\overline K^0 \omega$ amplitude, the
$\overline K^0 \omega/\pi^- \omega$ ratio should be very small.
Choosing smaller $m_s$ could enhance the ratio, but does not help
much in understanding data. We consider the contributions of the
three-parton Fock state of the final state $\pi$ (or $K$) meson to
the decay amplitudes, as shown in Fig.~\ref{fig:3p}. We find that
they can give significant corrections to decays with $\omega$ in
the final states. The decay amplitudes with corrections from the
three-parton Fock states are given by
 \bea
  &&{\cal A}( B^- \to \pi^-\omega)\nonumber\\
&&\ \ \ \simeq\cdots + G_F m_\omega (\epsilon^*_\omega \cdot
p_\pi)f_\omega F_1^{B\to \pi } (m_\omega^2) (V_{ub}V_{ud}^*c_1
-V_{tb}V_{td}^*
(2c_4-2c_6+c_3))f_3 \,,\nonumber\\
&&{\cal A}(\bar B^0 \to \bar K^0 \omega)\nonumber\\
 &&\ \ \ \simeq \cdots + G_F
m_\omega (\epsilon^*_\omega \cdot p_K ) f_\omega F_1^{B\to K
}(m_\omega^2) ( V_{ub}V_{us}^* c_1 - V_{tb}V_{ts}^* (2c_4
-2c_6))f_3,\eea where ``$\cdots$" denote the contributions from
two-parton Fock states of the final state mesons, the
normalization scale of $c_i$ is $\sim$ 1 GeV, and
 \bea f_3&=&
\frac{\sqrt{2}}{m_B^2 f_\omega F_1^{B\to \pi }(m_\omega^2)}\langle
\omega \pi^-|O_1|B^-\rangle_{\rm
qqg}\nonumber\\
&=&-\frac{4}{ \overline\alpha_g^\pi m_B^4 F_1^{B\to \pi
}(m_\omega^2) }p_{\omega}^\alpha \langle \pi^-|\bar d \gamma^\mu
\gamma_5 g_s \widetilde{G}_{\alpha\mu} b|B^-\rangle\label{eq:3p}
\simeq 0.12, \eea with $O_1=\overline d_\alpha \gamma^\mu
(1-\gamma_5)u_\alpha \cdot \overline u_\beta \gamma_\mu
(1-\gamma_5) b_\beta$, and $\overline\alpha_g^\pi\approx 0.23$
being the averaged fraction of the pion momentum carried by the
gluon. More details of the present work would be given
elsewhere~\cite{yang1}. We plot in Fig.~\ref{fig:omegakpi} the
branching ratios for $K\omega, \pi\omega$ vs the weak angle
$\gamma$. Taking $\gamma=90^\circ$, we obtain the branching ratios
$\pi^-\omega: K^-\omega:\overline K^0\omega\simeq 5.5: 4.5: 4.3$
in units of $10^{-6}$, which are in good agreement with the
present data.
%
\bfg[t]                     
\bc                         
\epsfig{figure=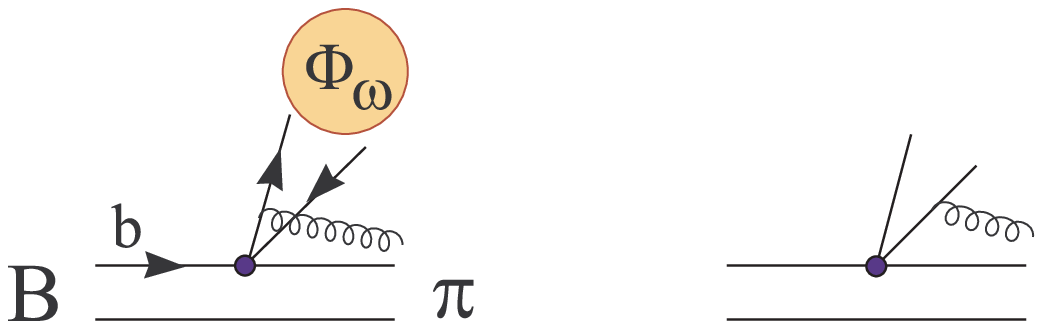,height=25mm}
\ec                         
\caption{The contributions of the $d \bar u g$ Fock state of the
pion to the $B^-\to \pi^- \omega$ amplitude.}\label{fig:3p} \efg

%
\bfg[t]                     
\bc                         
\epsfig{figure=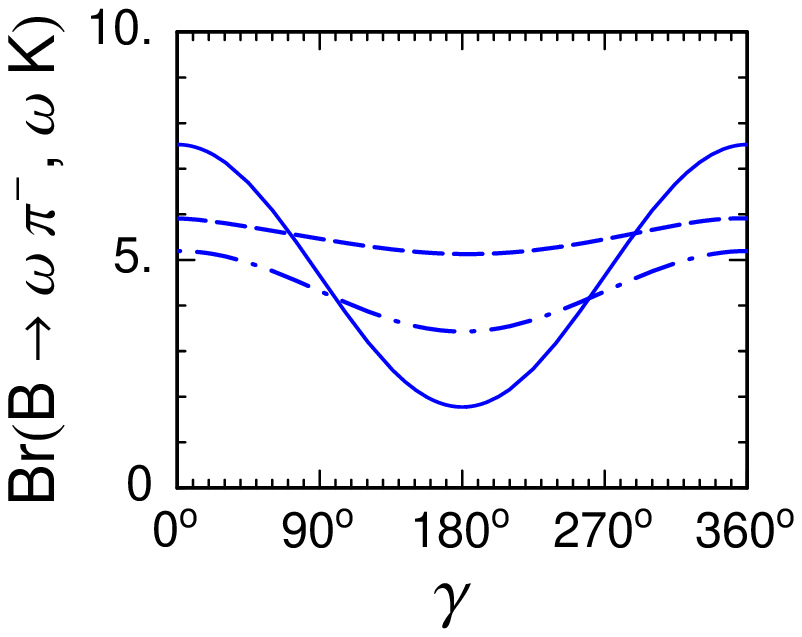,height=45mm}
\ec                         
\caption{Branching ratios for $\pi\omega$ and $K\omega$ vs
$\gamma$. The dashed, solid, and dot-dashed lines denote
$\overline B\to \pi^-\omega, K^-\omega$ and $\overline K^0\omega$,
respectively, with $\rho_A=0.9$ and $f_B=180$~MeV.
}\label{fig:omegakpi} \efg

\section{Summary}
Using the QCD factorization approach, we study various important
corrections from higher twist distribution amplitudes of mesons in
the hadronic $B$ decays.  (i) For the case of $B\to \phi K$, the
weak annihilation diagrams induced by $(S-P)(S+P)$ penguin
operators, which are power-suppressed by order of $(\Lambda_{\rm
QCD}/m_b)^2$, are chirally and in particular logarithmically
enhanced owing to the non-vanishing end-point behavior of the
twist-3 light-cone distribution amplitudes of the kaon. (ii) For
the case of $B\to VV$, it is necessary to take into account
twist-3 distribution amplitudes of the vector meson in order to
have renormalization scale and scheme independent predictions.
(iii) $a_2(J/\psi K)$ may be largely enhanced by the
nonfactorizable spectator interactions arising from the twist-3
kaon LCDA $\Phi^K_\sigma$, which are formally power-suppressed but
chirally and logarithmically enhanced. However, since the
contribution of the twist-3 kaon LCDA $\Phi^K_\sigma$ to the
spectator diagram is end-point divergent in the collinear
expansion, the vertex of the gluon and spectator quark should be
considered to be inside the kaon wave function. Instead of
considering the contribution of the two-parton kaon LCDA of
twist-3 to spectator diagrams, we calculate the subleading
corrections originating from the three-parton LCDAs of the kaon
and obtain $a_2= 0.27+0.05i$ which is well consistent with the
data and can solve the long-standing sign ambiguity of $a_2$. (iv)
We study the subleading corrections arising from the three-parton
Fock states of mesons in $B$ decays. Our results can account for
the observation of $\omega K \sim \omega \pi^-$.

\bigskip
{\small Acknowledgement: I wish to thank Hai-Yang Cheng for
collaboration. This work is supported in part by the National
Science Council of R.O.C. under Grant No. 91-2112-M-033-013.}

\end{document}